\magnification=1200
\settabs 18 \columns

\baselineskip=12 pt
\topinsert \vskip 0.75 in
\endinsert

\def\sqr#1#2{{\vcenter{\vbox{\hrule height.#2pt
 \hbox{\vrule width.#2pt height#1pt \kern#1pt
 \vrule width.#2pt} \hrule height.#2pt}}}}

\def\operp{\hbox{${\kern+.25em{\bigcirc}
\kern-.85em\bot\kern+.85em\kern-.25em}$}}

\def\lsim{\;\raise0.3ex\hbox{$<$\kern-0.75em\raise-1.1ex\hbox{$\sim$}}\;}
\def\gsim{\;\raise0.3ex\hbox{$>$\kern-0.75em\raise-1.1ex\hbox{$\sim$}}\;}
\def\no{\noindent}

\def\ce{\centerline}
\def\ve{\vfill\eject}
\def\rdots{\mathinner{\mkern1mu\raise1pt\vbox{\kern7pt\hbox{.}}\mkern2mu
 \raise4pt\hbox{.}\mkern2mu\raise7pt\hbox{.}\mkern1mu}}

\def\e e{$e^+ e^-$ }



\rightline{UCLA/01/TEP/12}
\rightline{\bf Revised Version}
\vskip1.5cm

\baselineskip=15pt

\ce{{\bf SOLITONIC ASPECTS OF $q$-FIELD THEORIES}}
\vskip.5cm
\ce{\it R. J. Finkelstein}
\vskip.3cm
\ce{Department of Physics and Astronomy}
\ce{University of California, Los Angeles, CA  90095-1547}
\vskip1.0cm

\no {\bf Abstract.}  We have examined the deformation of a generic
non-Abelian gauge theory obtained by replacing its Lie group by the
corresponding quantum group.  This deformed gauge theory has more degrees
of freedom than the theory from which it is derived.  By going over from
point particles in the standard theory to solitonic particles in the
deformed theory, it is proposed to interpret the new degrees of fredom as
descriptive of a non-locality of the deformed theory.  It also turns out that
the original Lie algebra gets replaced by two dual algebras, one of which
lies close to and approaches the original Lie algebra in a correspondence
limit, while the second algebra is new and disappears in this same
correspondence limit.  The exotic field particles associated with the second
algebra can be interpreted as quark-like constituents of the solitons, which
are themselves described as point particles in 
the first algebra.  These ideas are explored for
$q$-deformed $SU(2)$ and $GL_q(3)$.

\ve

\line{{\bf 1. Introduction.} \hfil}
\vskip.3cm

The quantum groups made their original appearance in connection with the
inverse scattering problem and Yang-Baxter physics.$^1$  Later they became
important in the theory of knots and in conformal field theory.  Since the
Lie groups may be regarded as degenerate forms of the quantum groups, it may
also be of interest to replace the latter by the quantum groups in other
physical contexts.  For example, corresponding to quantum mechanical
systems such as the harmonic oscillator and the hydrogen atom there are
$q$-systems obtained by going over to $q$-groups.  It is then found that the
quantum mechanical $q$-systems have more degrees of freedom than the systems
from which they are derived.  When field theories based on gauged Lie groups
are similarly deformed by replacing the Lie groups by the corresponding quantum
groups, the new degrees of freedom may be interpreted as an expression of
the non-locality exhibited by extended or solitonic particles.$^2$  Here we
explore some aspects of $q$-theories lying formally close to the standard
model.  It is proposed that the original Lie-based theory be replaced by a
$q$-theory with two sectors: one describing the point particles
representing the solitons,
that may approximate the standard theory, while the second sector describes the
dynamics of the fields comprising the solitons.

We first describe the simplest non-trivial quantum groups $SL_q(2),~SU_q(2)$
and the attached Hilbert space.
\vskip.5cm

\line{{\it Two-Dimensional Representation of $SL_q(2)$} \hfil}
$$
T\epsilon T^t = T^t\epsilon T = \epsilon \qquad
T\epsilon SL_q(2) \eqno(1.1)
$$
\no where $t$ means transpose and
$$
\epsilon = \left(\matrix{0 & q_1^{1/2} \cr -q^{1/2} & 0 \cr} \right)
\qquad q_1 = q^{-1} \eqno(1.2)
$$
\no Set
$$
T = \left(\matrix{\alpha & \beta \cr \gamma & \delta \cr} \right)
\eqno(1.3)
$$
\no Then
$$
\eqalign{&{\rm (a)}~~\alpha\beta = q\beta\alpha \cr
&{\rm (b)}~~\delta\beta = q_1\beta\delta \cr 
& \hfil \cr} \qquad
\eqalign{&{\rm (c)}~~\alpha\gamma = q\gamma\alpha \cr
&{\rm (d)}~~\delta\gamma = q_1\gamma\delta \cr
& \hfil \cr} \qquad
\eqalign{&{\rm (e)}~~\alpha\delta-q\beta\gamma = 1 \cr
&{\rm (f)}~~\delta\alpha-q_1\beta\gamma = 1 \cr
&{\rm (g)}~~\beta\gamma = \gamma\beta \cr} \eqno(1.4)
$$
\no If $q=1$, the equations (1.4) are satisfied by complex numbers, and
$T$ is defined over a continuum; but if $q\not= 1$, then $T$ is defined only
over the algebra--a non-commuting space.

We adopt a matrix representation of the algebra and impose $\delta = \bar\alpha$
and $\beta = \bar\beta$, $\bar\gamma = \gamma$ where the bar means Hermitian
conjugate.  There are no finite representations of this algebra unless $q$
is a root of unity.

To define the state space attached to the algebra we interpret $\bar\alpha$
and $\alpha$ as raising and lowering operators respectively as follows.
\vskip.5cm

\line{{\it Ground state} \hfil}
$$
\eqalignno{\alpha|0\rangle &= 0  & (1.5) \cr
\beta|0\rangle &= b|0\rangle & (1.6a) \cr
\gamma|0\rangle &= c|0\rangle & (1.6b) \cr}
$$
\vskip.5cm

\line{{\it Excited states} \hfil}
$$
\bar\alpha|n\rangle = \lambda_n|n+1\rangle \eqno(1.7)
$$
\no By iterating
$$
\eqalign{\beta\bar\alpha &= q\bar\alpha\beta \cr
\gamma\bar\alpha &= q\bar\alpha\gamma \cr} 
$$
\no we find
$$
\eqalignno{\beta|n\rangle &= q^nb|n\rangle & (1.8) \cr
\gamma|m\rangle &= q^mc|m\rangle & (1.9) \cr}
$$
\no By (1.4f)
$$
(\bar\alpha\alpha-q_1\beta\gamma)|0\rangle = |0\rangle \eqno(1.10)
$$
\no Hence
$$
bc = -q \eqno(1.11)
$$
\no Also by (1.4e)
$$
\langle n|n\rangle = 1\to \lambda_n = (1-q^{2n+2})^{1/2} \eqno(1.12)
$$
\no and
$$
q<1
$$
\ve

\line{{\it Two-Dimensional Representation of $SU_q(2)$} \hfil}
\vskip.3cm

Introduce matrix representation of algebra (1.4) and set
$$
\gamma = -q_1\bar\beta \qquad \delta = \bar\alpha \eqno(1.13)
$$
\no where bar means hermitian conjugate.  Then
$$
\eqalign{\alpha\beta &= q\beta\alpha \cr
\alpha\bar\beta &= q\bar\beta\alpha \cr} \quad
\eqalign{\alpha\bar\alpha &+ \beta\bar\beta = 1 \cr
\bar\alpha\alpha &+ q_1^2\bar\beta\beta = 1 \cr} \quad
\beta\bar\beta = \bar\beta\beta
\eqno(A)
$$
\no and
$$
T \quad \hbox{is unitary:}~~\bar T = T^{-1} \eqno(1.14)
$$
\no If $q=1$, $(A)$ may be satisfied by complex numbers and $T$ is the usual
$U(2)$ unitary-symplectic matrix.  If $q\not= 1$, there are no finite
representatiions of $(A)$ unless $q$ is a root of unity.  Therefore, as before
\item{} If $q=1$, $T$ is defined over a continuum.
\item{} if $|q| \not= 1$, $T$ is defined over algebra $(A)$-a non-commuting
space.

We shall assume that $q$ is real.

The irreducible representations of $SU_q(2)$ are as follows:
$$
\eqalign{
{\cal{D}}^j_{mm^\prime}(\alpha,\bar\alpha,\beta,\bar\beta) = \Delta^j_{mm^\prime}
&\sum_{s,t} \biggl\langle\matrix{n_+\cr s\cr}\biggr\rangle_1
\biggl\langle\matrix{n_-\cr t\cr}\biggr\rangle_1 q_1^{(n_1-s+1)t}
(-)^t \cr 
&\times\delta(s+t,n_+^\prime)\alpha^s\beta^{n_+-s}
\bar\beta^t\bar\alpha^{n_--t} \cr} \eqno(1.15)
$$
\no where
$$
\eqalign{n_\pm &= j\pm m \cr
n_\pm^\prime &= j\pm m^\prime \cr} \qquad
\biggl\langle\matrix{n\cr s \cr}\biggr\rangle_1 = {\langle n\rangle_1!\over
\langle s\rangle_1! \langle n-s\rangle_1!} \qquad
\langle n\rangle_1 = {q_1^{2n}-1\over q_1^2-1}  \eqno(1.16)
$$
$$
q_1 = q^{-1} \qquad
\Delta^j_{mm^\prime} = \biggl[{\langle n_+^\prime\rangle!
\langle n^\prime_-\rangle!\over \langle n_+\rangle!
\langle n_-\rangle!}\biggr]^{1/2} \eqno(1.17)
$$
\no In the limit $q=1$ the ${\cal{D}}^j_{mm^\prime}$ become the Wigner
functions, $D^j_{mm^\prime}(\alpha,\beta,\gamma)$, the irreducible
representations of $SU(2)$.  The orthogonality relations may be expressed
in the following way
$$
\eqalignno{\int_{SU_q(2)} &\bar{\cal{D}}^j_{m\mu}(\alpha,\bar\alpha,
\beta,\bar\beta){\cal{D}}^{j'}_{m'\mu'}(\alpha,\bar\alpha,\beta,\bar\beta)
d\tau(\alpha,\bar\alpha,\beta,\bar\beta) = 
{\delta^{jj'}\delta_{mm'}\delta_{\mu\mu'}\over
[2j+1]_q} q^{2\mu} & (1.18) \cr
&[2j+1]_q = {q^{2j+1}-q_1^{2j+1}\over q-q_1} & (1.19) \cr}
$$
\no where the integral means an integral over the algebra and should be
understood in terms of the Haar measure as defined by Woronowicz.$^3$

We shall comment on the $q$-Lorentz group in the next section but the rest
of the paper deals with the $q$-gauge group.

In Sections III-V it is shown how point particles are replaced by solitons
if the field operator lies in the algebra of a $q$-global gauge group.

In Section VI the extension to a local gauge group is described.

Sections VII and VIII deal with two options for the vector connection, one
lying in the $q$-algebra and the other in the dual algebra.

While III-VIII are based on $SU_q(2)$ and $SL_q(2)$, Sections IX-XII deal
with $GL_q(3)$ and $SL_q(3)$.

Finally XIV describes a toy field.
\vskip.5cm

\line{{\bf II. $q$-Fields.} \hfil}
\vskip.3cm

The structure of the standard field theories is determined by both the Lorentz
group and the gauge group.  Consider first the Lorentz group $SL(2,C)$. 
Replace $SL(2,C)$ by $SL_q(2,C)$.  In the spin representation one has
$$
\eqalignno{{\rm Lorentz:} \quad &\epsilon = L^t\epsilon L \qquad
\epsilon = \left(\matrix{0 & 1 \cr -1 & 0 \cr} \right) & (2.1) \cr
{\hbox{$q$-Lorentz:}} \quad &\epsilon_q = L^t\epsilon_qL \qquad
\epsilon_q = \left(\matrix{0 & q_1^{1/2} \cr -q^{1/2} & 0 \cr} \right) &
(2.2) \cr}
$$
Since Pauli has shown that the Lorentz group establishes a connection
between spin and statistics, one might expect this connection to be
changed when one goes over to the $q$-Lorentz group.
To illustrate this question let us consider a vector-spinor field with the
conventional Lorentz invariant interaction
$$
\bar\psi A\!\!\!/\psi
$$
\no and let us pursue conventional theory aside from the introduction of
the following time ordered products:
$$
\eqalign{T_q(\psi(x)\psi(x^\prime)) &=\psi(x)\psi(x^\prime) \qquad
t>t^\prime \cr
&= q^\prime\psi(x^\prime)\psi(x) \quad t<t^\prime~~\hbox{vector} \cr
&= q^{\prime\prime}\psi(x^\prime)\psi(x) \quad t<t^\prime~~\hbox{spinor}
\cr}
$$

Then the $q$-time-ordered $S$-matrix is
$$
S^{(q)} = T_q\bigl(e^{i\int {\cal{L}}(x)d^4x}\bigr)
$$
\no where $(q) = (q^\prime,q^{\prime\prime})$, and the Wick expansion
leads to Feynman rules with the following $q$-propagators$^4$
$$
\eqalign{D_{\mu\lambda}^{q^\prime}(x) &= \biggl(g_{\mu\lambda} -
{\partial_\mu\partial_\lambda\over m^2}\biggr)\biggl({1\over 2\pi}\biggr)^4
\biggl({1+q^\prime\over 2}\biggr) \int e^{-ikx}
{1\over k^2-m^2}\biggl[1 + {1-q^\prime\over 1+q^\prime}{k_o\over\omega}
\biggr] d^4k \cr
S_{\alpha\beta}^{q''}(x) &= (\partial\!\!\!/+m)_{\alpha\beta}
\biggl({1\over 2\pi}\biggr)^4\biggl({1-q^{\prime\prime}\over 2}\biggr)
\int e^{-ikx}{1\over k^2-m^2}
\biggl[1 + {1+q^{\prime\prime}\over 1-q^{\prime\prime}}{k_o\over\omega}
\biggr] d^4k~. \cr}
$$
\no The difference between these two propagators comes from $(q^\prime,q^{\prime\prime})$ and the sum over polarization and antiparticle
states.  The final theory may be tested for Lorentz invariance by calculating particle-particle scattering which depends on the photon
propagator, and particle-antiparticle annihilation, which depends on the
spinor propagator.

In both cases the result is frame dependent, i.e., Lorentz invariance is
broken unless $q^\prime=1$ for the vector and $q^{\prime\prime} = -1$ for the spinor propagator.  This result is then a special case of the Pauli
theorem that Lorentz invariance requires commutation and anticommutation
rules for integer and half-integer spin respectively.

Our question is now: if the $L$ group is replaced by $L_q$, are $q^\prime$ and $q^{\prime\prime}$ still constrained to be +1 and -1 or will one
find two new functions: $q^\prime=q^\prime(q)$ and $q^{\prime\prime} =
q^{\prime\prime}(q)$?

The answer to this question is that one does not find two new functions:
it is still possible to require TCP and the usual connection between spin
and statistics$^5$ with $L_q$.

If the $q$-Lorentz group is gauged, one obtains $q$-gravity.

We do not discuss the $q$-Lorentz group further.  The remainder of this
paper is devoted to the $q$-gauge groups.
\vskip.5cm

\line{{\bf III.  $q$-Scalar Theory.$^2$} \hfil}
\vskip.3cm

While retaining Lorentz invariance, we shall now examine a scalar theory
invariant under global $SL_q(2)$.
\vskip.3cm

\line{{\it Classical Hamiltonian:} \hfil}
$$
\eqalignno{H &= {1\over 2}\int\bigl[\vec\pi^2 + (\vec\nabla\psi)^2 + m^2\psi^2\bigr] d\vec x \cr
\noalign{\hbox{or}}
H &= {1\over 2}\int \bigl[\sum^3_{k=0}(\partial_k\psi)(\partial_k\psi)
+m^2\psi^2\bigr] d\vec x & (3.1) \cr}
$$

\ve

\line{{\it Hamiltonian Invariant under $SL_q(2)$:} \hfil}
\vskip.3cm
$$
H^{(q)} = {1\over 2}\int\bigl[\sum^3_{k=0}\partial_k\tilde\psi\epsilon\partial_k
\psi + m^2\tilde\psi\epsilon\psi\bigr]d\vec x \eqno(3.2)
$$
\no $H^{(q)}$ is invariant under
$$
\eqalign{&\psi^\prime = T\psi \qquad
(T^t\epsilon T=\epsilon) \cr 
&\tilde\psi^\prime = \tilde\psi T^t \cr} \eqno(3.3)
$$

Now $\psi$ must lie in the $q$-algebra.
\vskip.3cm

\line{{\it Quantization of Scalar Field Operator:} \hfil}
\vskip.3cm
$$
\eqalign{\psi(x) = &\biggl({1\over 2\pi}\biggr)^{3/2}\int {d\vec p\over
(2p_o)^{1/2}}\bigl(\sum_s u(s)\tau(s)\bigr) \cr
&\times \bigl[e^{-i\vec p\vec x}
a(\vec p) + e^{i\vec p\vec x}\bar a(\vec p)\bigr] \cr} \eqno(3.4)
$$
\no where we shall now assume
$$
\eqalignno{&\tau(s)=(\beta,\gamma) & (3.5) \cr
&\tilde u(s)\epsilon u(s^\prime) = \delta(s,s^\prime) U(s) & (3.6) \cr
&(a(\vec p),\bar a(\vec p^{~\prime})) = \delta(\vec p-\vec p^{~\prime}) 
& (3.7) \cr}
$$

\line{{\it Expanded State Space:} \hfil}
$$
\eqalign{|N(\vec p)n_1n_2\rangle 
\sim ~&\hbox{state of $N$ particles, each with} \cr
& \hbox{momentum $\vec p$ and quantum numbers} \cr
&\hbox{$(n_1n_2)$ belonging to $(\beta,\gamma)$} \cr}  
$$

\no The $|N(\vec p)n_1n_2\rangle$ are generated by application of raising
operators $\bar a(\vec p)$ and $\bar\alpha$.
\vskip.5cm

\line{{\bf IV. Reduction of Quantized Hamiltonian} \hfil}
\vskip.3cm

Assume normal ordering.  Then by (3.2), (3.4) and (3.5)
$$
\eqalignno{H^q &= {1\over 2} \int :\bigl[\sum^3_0 \partial_k\tilde\psi
\epsilon\partial_k\psi + m^2\tilde\psi\epsilon\psi\bigr]: d\vec x & (4.1) \cr
&=\biggl({1\over 2\pi}\biggr)^3 \int d\vec x \int
{d\vec p d\vec p^{~\prime}\over 2(p_op^\prime_o)^{1/2}}
\bigl[\sum^3_0 p_kp_k^\prime+m^2\bigr] & (4.2) \cr
& ~~~\times e^{i(p-p^\prime)x}[\tilde U\epsilon U]
:{1\over 2}\bigl[\bar a(p)a(p^\prime)+a(p^\prime)\bar a(p)\bigr]: & (4.3)\cr}
$$
\no where
$$
\eqalignno{[\tilde U\epsilon U] &= \sum_{ss^\prime}
\bigl(\tau(s)\tilde u(s)\bigr)\epsilon\bigl(u(s^\prime)\tau(s^\prime)\bigr)
& (4.4) \cr
&= \sum_s U(s)\tau(s)^2 & (4.5) \cr}
$$
\no by (3.6).  Then
$$
\eqalignno{H^q &= \int {d\vec p\over 2p_o}\bigl[\sum^3_0 p_k^2+m^2\bigr]
\bar a(p) a(p) [\tilde U\epsilon U] & (4.6) \cr
\noalign{\hbox{or}}
H^q &= \int d\vec pp_o(\tilde U\epsilon U)\bar a(p)a(p) & (4.7) \cr
\noalign{\hbox{and}}
H^q|N_pn_1n_2\rangle &= \int d\vec pp_o N_p(\tilde U\epsilon U)|N_pn_1n_2\rangle & (4.8) \cr}
$$
\no By (4.5) and (3.5)
$$
(\tilde U\epsilon U)|N_pn_1n_2\rangle = [U(1)\beta^2 + U(2)\gamma^2]|N_pn_1n_2\rangle \eqno(4.9)
$$
\no By (1.8) and (1.9)
$$
H^q|N_pn_1n_2\rangle = \int d\vec pp_oN_p[U(1)q^{2n_1}b^2 + U(2)
q^{2n_2}c^2]|N_pn_1n_2\rangle \eqno(4.10)
$$
\no Then we find the energy of a single particle:
$$
p_o[U(1)q^{2n_1}b^2 + U(2)q^{2n_2}c^2] \eqno(4.11)
$$
\no Set $\vec p=0$, then the mass of a single particle is by (1.11)
$$
m(n_1,n_2) = m\biggl[U(1)q^{2n_1}b^2 + U(2) q^{2n_2+2}{1\over b^2}\biggr]
\eqno(4.12)
$$
\no where $U(i) = \tilde u(i)\epsilon u(i),~i=1,2$.  Now introduce the
length, $R$, by
$$
U(1)b^2 = {1\over R} \eqno(4.13)
$$
\no Then
$$
m(n_1,n_2) = m\biggl[q^{2n_1}{1\over R} + q^{2n_2+2}(U(1)U(2)R\biggr] 
\eqno(4.14)
$$
\no This spectrum resembles the spectrum of a toroidally compactified string
with an associated large-small $(T)$ duality:$^4$ 
$$
R\to R^\prime = (q^2U(1)U(2)R)^{-1} \qquad n_1\leftrightarrow
n_2 \eqno(4.15)
$$
\no It is self-dual with the characteristic length
$$
\tilde R = (q^2U(1)U(2))^{-1/2} \quad \hbox{and} \quad
n_1=n_2 \eqno(4.16)
$$
\vskip.3cm

\line{{\it Remarks} \hfil}
\vskip.3cm

Since $q<1$, the highest mass is $m$.  If $\alpha$ and $\bar\alpha$ are interchanged in $(A)$, then $q\to q_1$ in (4.14) and the spectrum is
inverted.

The point particles of the field theory with $q=1$ have now become solitons
since
any kind of a mass spectrum must be interpreted to imply extension in 
spacetime.  Assuming that the mass of a field quantum is $\sim q^{2n}$, one
may gain some idea of its spatial extension by noting that $q^{2n}$
resembles the spectrum $(\sim\langle n\rangle_{q^2})$ of a 
$q$-harmonic oscillator. The wave functions of a $q$-oscillator
are $q$-Hermite functions.$^6$  A similar shape may be associated with the field
soliton while the size may be related to the characteristic length 
$(q^2U(1)U(2))^{-1/2}$.

If the scalar field is charged, one finds that the charge and mass of the
soliton are proportional (as in BPS solitons).
\vskip.5cm

\line{{\bf V. A General Gauge.} \hfil}
\vskip.3cm

The equations (3.4) and (3.5) may be regarded as defining a special
gauge, according to which the field operator $\psi$ lies in the
$\beta,\gamma$ subspace.  More generally we may replace (3.4) by an
expansion in the irreducible representations of $SU_q(2)$
$$
\eqalign{\psi(x) &= u\sum_{jmn} \varphi^j_{mn}(x) {\cal{D}}^j_{mn}(\alpha|q) \cr
\tilde\psi(x) &= \sum_{jmn} \tilde\varphi^j_{mn}(x)
\bar{\cal{D}}^j_{mn}(\alpha|q)u^t \cr} \eqno(5.1)
$$
\no where we have also substituted $SU_q(2)$ for $SL_q(2)$.  Here
complex conjugation is denoted by a bar.  We suppose that
$\varphi^j_{mn}(x)$ does not lie in the $q$-algebra and that $u$ is
a 2-rowed vector transforming as
$$
\eqalign{u^\prime &= Tu \cr
u^{t'} &= u^tT^t \cr} \eqno(5.2a)
$$
\no and normalized according to
$$
u^t\epsilon u = 1 \eqno(5.2b)
$$

The partial fields $\varphi^j_{mn}(x)$ appearing in (5.1) may themselves
be expanded in terms of annihilation and creation operators
$$
\varphi^j_{mn}(x) = {1\over (2\pi)^{3/2}}\int {d\vec p\over (2p_o)^{1/2}}
\bigl[e^{-ipx}a^j_{mn}(\vec p) + e^{ipx}\bar a^j_{mn}(\vec p)\bigr]~.
\eqno(5.3)
$$
\no Here $a^j_{mn}(\vec p)$ is the absorption operator for a particle of
momentum $\vec p$ and additional quantum numbers $(jmn)$.  Instead of
evaluating the Hamiltonian over a particular state we now average over the
full algebra.  We therefore replace (3.2) by
$$
H^q = {1\over 2} h\int : \biggl[\sum_0^3\partial_k\tilde\psi\epsilon
\partial_k\psi + m_o^2\tilde\psi\epsilon\psi\biggr]:d\vec x \eqno(5.4)
$$
\no where the symbol $h$ standing before the spatial integral denotes
a Woronowicz integral over the $SU_q(2)$ algebra.  The evaluation of
(5.4) with (5.1) and (5.3) gives
$$
H^q = \int d\vec pp_o \sum_{\scriptstyle jmn\atop\scriptstyle j^\prime m^\prime n^\prime} h\bigl(\bar{\cal{D}}^j_{mn}{\cal{D}}^{j'}_{m^\prime n^\prime}\bigr)
{1\over 2}:\bigl[\bar a^j_{mn}(\vec p)a^{j'}_{m^\prime n^\prime}(\vec p)
+ a^{j^\prime}_{m^\prime n^\prime}(\vec p)
\bar a^j_{mn}(\vec p)\bigr]: \eqno(5.5)
$$

The orthogonality of the $q$-irreducible representations is now expressed
in terms of the Haar measure as in (1.18)
$$
h(\bar{\cal{D}}^j_{mn}{\cal{D}}^{j^\prime}_{m^\prime n^\prime}) = \delta^{jj^\prime}
\delta_{mm^\prime}\delta_{nn^\prime} {q^{2n}\over [2j+1]_q}~.
\eqno(5.6)
$$
\no Then
$$
H^q|N(p);jmn\rangle = {p_oq^{2n}\over [2j+1]_q}
|N(\vec p);jmn \rangle \eqno(5.7)
$$
\no Therefore the rest mass of a single particle with ``internal" quantum
numbers $(jmn)$ is
$$
{m_oq^{2n}\over [2j+1]_q}~. \eqno(5.8)
$$
\no This spectrum resembles the square root of the spectrum of the
$q$-H atom.$^7$ 

In the $\beta\gamma$-gauge we evaluated the rest mass in a particular
state of expanded state space.  In the more general expansion above we
chose to average over the full state space.  The $\varphi^j_{mn}(x)$ may be
regarded as constituent or preon fields.
\ve

\line{{\it Spinor Case} \hfil}

If $\psi$ is a Dirac spinor, the usual mass term is
$$
M\psi^t C\psi \eqno(5.9)
$$
\no invariant under Lorentz transformations, $L$, since
$$
L^tCL = LCL^t = C \eqno(5.10)
$$
\no where $C$ is the charge conjugation matrix.

Invariance under independent $L$ and $T$ transformations requires
$$
M\tilde\psi C\epsilon\psi \eqno(5.11)
$$
\no Then the spinor solitons would have mass $\sim Mq^n$.

(If Lorentz symmetry is broken, one possibility is the replacement of $L$ by
$L_q$ and $C$ by $\epsilon$.  Then the term invariant under $L_q$ is
$$
M\tilde\psi\epsilon\psi \eqno(5.12)
$$
\no All of the usual bilinear covariants as well as
the relation between particle and antiparticle would then depend on $q$
when $C$ is replaced by $\epsilon$.)
\vskip.5cm 

\line{{\bf VI. Local Theory.} \hfil}
\vskip.3cm

We assume that $T$ is position dependent as in a Yang-Mills theory.  Then
$$
T(x)^t\epsilon T(x) = T(x)\epsilon T^t(x) = \epsilon \eqno(6.1)
$$

\line{{\it Mass Terms:} \hfil}
$$
\eqalign{
&{\cal{M}}_1\tilde\psi C\epsilon\psi + {\cal{M}}_2^2\tilde\varphi\epsilon\varphi\cr
&~\scriptscriptstyle{\hbox{spinor}} \qquad ~\scriptscriptstyle{\hbox{scalar}} \cr} \eqno(6.2)
$$
\no where
$$
\eqalign{\psi^\prime &= T\psi \cr
\tilde\psi^\prime &= \tilde\psi T^t \cr} \qquad
\eqalign{\varphi^\prime &= T\varphi \cr
\tilde\varphi^\prime &= \tilde\varphi T^t \cr} \eqno(6.3)
$$
\vskip.3cm

\line{{\it Kinetic Terms:} \hfil}
$$
\eqalign{\tilde\psi & C\epsilon\gamma^\mu\vec\nabla_\mu\psi + {1\over 2}
(\tilde\varphi\buildrel\leftarrow\over{\nabla}_\mu)\epsilon(\vec\nabla^\mu
\varphi) \cr
&~\scriptscriptstyle{\hbox{spinor}} \qquad ~~~~\quad \scriptscriptstyle{\hbox{scalar}}
\cr} \eqno(6.4)
$$
\no where $\vec\nabla_\mu$ and $\buildrel\leftarrow\over{\nabla}_\mu$
are covariant derivatives.
\ve

\line{{\it Transformation Rules:} \hfil}
$$
\eqalign{\vec\nabla_\mu^\prime &= T\vec\nabla_\mu T^{-1} \cr
\vec\nabla_\mu &= \vec\partial_\mu + \vec A_\mu \cr} \qquad
\eqalign{\buildrel\leftarrow\over\nabla\!{}^\prime_\mu &= (T^t)^{-1}
\buildrel\leftarrow\over{\nabla}_\mu T^t \cr
\buildrel\leftarrow\over{\nabla}_\mu &= \buildrel\leftarrow\over{\partial}_\mu
+\buildrel\leftarrow\over{A}_\mu \cr} \eqno(6.5)
$$
$$
\eqalign{
\vec A^\prime_\mu &= T\vec A_\mu T^{-1}+T\vec\partial_\mu T^{-1} \cr
\buildrel\leftarrow\over{A}^\prime_\mu &=
(T^t)^{-1}\buildrel\leftarrow\over{A}_\mu T^t+(T^t)^{-1}
\buildrel\leftarrow\over{\partial}_\mu T^t \cr} \qquad
\eqalign{ &\hfil \cr
&\vec A_\mu = \epsilon\buildrel\leftarrow\over{A}_\mu\epsilon \cr}
\eqno(6.6)
$$
\no There is now a left as well as a right covariant derivative
$(\buildrel\leftarrow\over\nabla$ and $\vec\nabla$) and corresponding vector
connections ($\buildrel\leftarrow\over{A}_\mu$ and $\vec A_\mu$).
\vskip.5cm

\line{{\it Left and Right Field Strengths:} \hfil}
$$
\eqalign{\vec F_{\mu\nu} &= (\vec\nabla_\mu,\vec\nabla_\nu) \cr
\buildrel\leftarrow\over{F}_{\mu\nu} &= 
(\buildrel\leftarrow\over{\nabla}_\mu,\buildrel\leftarrow\over{\nabla}_\nu) \cr}
\quad
\eqalign{\vec F^\prime_{\mu\nu} &= T\vec F_{\mu\nu}T^{-1} \cr
\buildrel\leftarrow\over{F}^\prime_{\mu\nu} &= (T^t)^{-1}
\buildrel\leftarrow\over{F}_{\mu\nu}T^t \cr} \eqno(6.7)
$$
\no For the corresponding field invariants one may choose terms like
$$
\vec\varphi_\ell\vec F^{\mu\nu}\vec F_{\mu\nu}\vec\varphi_r \qquad
\buildrel\leftarrow\over{\varphi}_\ell
\buildrel\leftarrow\over{F}_{\mu\nu}\buildrel\leftarrow\over F\!{}^{\mu\nu}
\buildrel\leftarrow\over{\varphi}_r \eqno(6.8)
$$
\no where
$$
\eqalign{\vec\varphi_\ell^\prime &= \vec\varphi_\ell T^{-1} \cr
\buildrel\leftarrow\over\varphi\!{}^\prime_\ell &=
\buildrel\leftarrow\over{\varphi}_\ell T^t \cr} \qquad
\eqalign{\vec\varphi_r^\prime &= T\vec\varphi_r \cr
\buildrel\leftarrow\over\varphi\!{}^\prime_r &= (T^t)^{-1}
\buildrel\leftarrow\over{\varphi}_r \cr} \eqno(6.9)
$$
\no The usual trace Tr $F^{\mu\nu}F_{\mu\nu}$ is not invariant because
$$
(T_{ij},(F_{\mu\nu})_{k\ell}) \not= 0~. \eqno(6.10)
$$
\vskip.5cm

\line{{\bf VII. Representations of the Free Fields.} \hfil}
\vskip.3cm

The scalar and spinor expansions are simple extensions of the usual expressions but there are new options for the vector.
\vskip.3cm

\line{{\it scalar:} \hfil}
$$
\varphi = \biggl({1\over 2\pi}\biggr)^{3/2}\int {d\vec p\over (2p_o)^{1/2}}
\sum_s u(p,s)\bigl[e^{-ipx}a(p)+e^{ipx}\bar a(p)\bigr]\tau_s \eqno(7.1)
$$
\ve

\line{{\it spinor:} \hfil}
$$
\eqalign{\psi = \biggl({1\over 2\pi}\biggr)^{3/2}
\int &{d\vec p\over (2p_o)^{1/2}}
\sum_{r,s}\bigl[u(p,r,s) e^{-ipx}a(p,r) \cr
&+ v(p,r,s)e^{ipx}
\bar b(p,r)\bigr]\tau_s \cr} \eqno(7.2)
$$
\no where $\tau_s$ lies in $q$-algebra.
\vskip.3cm

\line{{\it vector:} \hfil}
\vskip.3cm

In the standard $SU(2)$ theory
$$
W_\mu = W_\mu(+)\tau(-) + W_\mu(-)\tau(+) + W_\mu(3)\tau_3 \eqno(7.3)
$$
\no In the $SU_q(2)$ theory one option is based on the following correspondence
between the $q$-algebra and the Cartan algebra of $SU(2)$:
$$
\bar\alpha\sim E_+ \quad\alpha\sim E_- \quad \hbox{and}
\quad (\beta,\gamma)\sim H \eqno(7.4)
$$
\no Then one may propose for the vector field lying in the $q$-algebra:
$$
A_\mu^{(q)} = A_\mu(+)\alpha + A_\mu(-)\bar\alpha +
A^{(\beta)}_\mu\beta + A_\mu^{(\gamma)}\gamma \eqno(7.5)
$$
\no There is a second option for the vector field, namely
$$
W_\mu^{(q)} = W_\mu(+) J^q(-) + W_\mu(-) J^q(+) +
W_\mu(3) J_3^q \eqno(7.6)
$$
\no based on the deformation of the Lie algebra where
$$
\eqalignno{&(J_3^{q},J^{q}_+) = J^{q}_+ \quad (J^{q}_3,J^{q}_-) = -J^{q}_- 
& (7.7) \cr
&(J^{q}_+,J^{q}_-) = {1\over 2}[2J^{q}_3]_q  & (7.8)\cr}
$$
\no Here
$$
[2J_3]_q = {q^{2J_3}-q^{-2J_3}\over q-q^{-1}} \qquad
\bigl(=q^{-2J_3+1}~~\langle 2J_3\rangle_{q^2}\bigr) \eqno(7.9)
$$

The first option is based on a deformation of the Lie group and the
second option is based on a deformation of the Lie algebra of the
same group.

Therefore there appear to be two sectors of the $q$-theory: one 
describing particles lying in the deformed group and the other describing
particles lying in the deformed algebra and suggesting 
that both $W_\mu^{(q)}$ and
$A_\mu^{(q)}$ should be included in the $q$-theory.
\vskip.5cm

\line{{\bf VIII. The Dual Algebras.} \hfil}
\vskip.3cm

The $q$-deformation of the Lie algebra, shown in (7.7)-(7.9), may be
obtained in the following way.

The two-dimensional representation, $T$, may be Borel factored:
$$
T = \left(\matrix{\alpha & \beta \cr \gamma & \delta \cr} \right) =
e^{B\sigma_+} e^{\lambda\theta\sigma_3} e^{C\sigma_-}~. \eqno(8.1)
$$
\no The algebra of $(\alpha,\beta,\gamma,\delta)$ is then inherited by
$(B,C,\theta)$ as
$$
(B,C) = 0 \qquad (\theta,B) = B \qquad (\theta,C) = C \qquad
\lambda = \ln q \eqno(8.2)
$$
\no The $2j+1$ dimensional irreducible representation of $SU_q(2)$ shown
in (1.15) may by (8.1) be rewritten in terms of $(B,C,\theta)$.  Then
one has by expanding ${\cal{D}}^j_{mm'}(B,C,\theta)$ to terms linear
in $(B,C,\theta)$
$$
\eqalign{{\cal{D}}^j_{mm'}(B,C,\theta) = {\cal{D}}^j_{mm'}(0,0,0) +
B(J^j_B)_{mm'} &+ C(J_C^j)_{mm'} + 2\lambda\theta(J_\theta^j)_{mm'} \cr
&+ \ldots  \cr} \eqno(8.3)
$$
\no where the non-vanishing matrix coefficients $(J_B^j)_{mm'}$,
$(J_C^j)_{mm'}$ and $(J_\theta^j)_{mm'}$ are
$$
\eqalign{\langle m-1|J_B^j|m\rangle &= \bigl[\langle j+m\rangle_1
\langle j-m+1\rangle_1\bigr]^{1/2} \cr
\langle m+1|J_C^j|m\rangle &=\bigl[\langle j-m\rangle_1
\langle j+m+1\rangle_1\bigr]^{1/2} \cr
\langle m|J_\theta^j|m\rangle &= m \cr} \eqno(8.4a)
$$
\no where
$$
\langle n\rangle_1 = {q_1^{2n}-1\over q_1^2-1} \eqno(8.4b)
$$
\no The $(B,C,\theta)$ and $(J_B,J_C,J_\theta)$ are generators of two
dual algebras satisfying the following commutation rules:
$$
\eqalignno{&(J_B,J_\theta) = -J_B \qquad (J_C,J_\theta) = J_C \qquad
(J_B,J_C) = q_1^{2j-1} [2J_\theta] & (8.5a) \cr
&(B,C) = 0 \qquad \qquad (\theta,B) = B ~\quad \qquad
(\theta,C) = C & (8.5b) \cr}
$$
\no We shall suppose that the $q$-theory describing deformations
($G_q$ and $g_q$) of both the group $G$ and the algebra $g$ is composed
of two sectors.  The $G_q$-sector contains new particles lying in the
algebra of $(\alpha,\bar\alpha,\beta,\gamma)$.  The $g_q$-sector lies
close to the standard theory and should approach the standard theory
in a $q=1$ correspondence limit.  In the soliton constructions the
$g_q$ particles are the solitons while the $G_q$ particles are related
to the constituent fields.  In the $q=1$ limit the solitons and the
$G_q$ particles vanish and the point particle picture is restored.
\vskip.5cm

\line{{\bf IX. Extension to Groups of Higher Rank.$^{8,9,10}$} \hfil}
\vskip.3cm

Although the standard theory is based on $U(1)\times SU(2)\times SU(3)$ it is
likely that it will ultimately be extended to embrace Lie groups of higher
rank.  We shall therefore briefly discuss the extension of
$SU_q(2)$ to $q$-groups of higher rank and in particular $GL_q(3)$.

In general the quantum groups may be defined by the relations$^8$
$$
RT_1T_2 = T_2T_1R \eqno(9.1)
$$
\no where
$$
\eqalign{T_1 &= T\otimes I \cr
T_2 &= I\otimes T \cr} \eqno(9.2)
$$
\no and the $R$-matrix satisfies the Yang-Baxter equation
$$R_{12}R_{13}R_{23} = R_{23}R_{13}R_{12}~.
\eqno(9.3)
$$
\no For the series $GL_q(N)^{~6}$
$$
R = q\sum^N_{i=1} e_{ii}\otimes e_{ii} + \sum^N_{\scriptstyle i,j=1\atop
\scriptstyle i\not= j} e_{ii}\otimes e_{jj} + (q-q_1)
\sum^N_{\scriptstyle i,j=1\atop\scriptstyle i>j} e_{ij}\otimes e_{ji}
\eqno(9.4)
$$
\no where
$$
[e_{mn}]_{ij} = \delta_{mi}\delta_{nj}~. \eqno(9.5)
$$
\no Solution of (9.1) with (9.4) leads to simple rules which replicate
the rules for $SL_q(2)$ as follows.  Let
$$
T = \left(\matrix{t_{11} & \ldots & t_{1N} \cr
\hfil & \hfil & \hfil \cr
t_{N1} & \ldots & t_{NN} \cr} \right) \qquad
T \epsilon GL_q(N) \eqno(9.6a)
$$
\no and let
$$
\left(\matrix{ik & i\ell \cr \hfil & \hfil \cr jk & j\ell \cr} \right)
\eqno(9.6b)
$$
\no be any rectangle of elements within $T$.  Then
$$
\eqalign{&t_{ik}t_{i\ell} = qt_{i\ell}t_{ik} \cr
&t_{ik}t_{jk} = qt_{jk}t_{ik} \cr
&(t_{ik},t_{j\ell}) = (q-q_1)t_{jk}t_{i\ell} \cr
&(t_{i\ell},t_{jk}) = 0 \cr} \eqno(9.7)
$$
\no i.e. the 4 vertices exhibited in (11.6b) belong to the algebra of
$GL_q(2)$.  Therefore all the commuting elements lie on lines of positive
slope and the maximum set of commuting elements, fixing the rank, lie on the
minor diagonal.

There is also the quantum determinant
$$
{\rm det}_qT = \sum_\sigma (-q)^{\ell(\sigma)}t_{1\sigma_1}\ldots
t_{N\sigma_N} \qquad \sigma~\epsilon~\hbox{symm}(N) \eqno(9.8)
$$
\no where $\ell(\sigma)$ is the number of inversions in going from the
order $1\ldots N$ to the permutation $\sigma$.

The quantum determinant commutes with all the elements of $T$
$$
(\hbox{det}_qT,t_{ij}) = 0 \eqno(9.9)
$$
\vskip.5cm

\line{{\bf X.  The Quantum Group $GL_q(3)$} \hfil}
\vskip.3cm

In a matrix realization of the algebra (9.7) the following structure is
permitted if $N=3$.
$$
T = \left(\matrix{\bar E_1 & \bar E_2 & H_a \cr
\bar E_3 & H_b & E_2 \cr H_c & E_3 & E_1 \cr} \right) \eqno(10.1)
$$
\no where the bar means Hermitian conjugation and
$$
\eqalign{H_i &= \bar H_i \cr q &= \bar q \cr} \qquad
i = a,b,c \eqno(10.2)
$$
\no The full set of commutation relations up to their hermitian
conjugates follows:
$$
\eqalign{{\rm a)}~&H_aE_2 = qE_2H_a \cr
&H_bE_3 = qE_3H_b \cr &H_cE_1 = qE_1H_c \cr} ~~
\eqalign{{\rm b)}~&H_aE_1 = qE_1H_a \cr
&H_cE_3 = qE_3H_c \cr &H_bE_2 = qE_2H_b \cr} ~~
\eqalign{{\rm c)}~&(H_a,E_3) = 0 \cr
&(H_c,E_2) = 0 \cr &(H_a,H_b) = (H_b,H_c) = (H_a,H_c) = 0 \cr}
\eqno(10.3)
$$
$$
\eqalign{{\rm a)}~(\bar E_1,E_1) &= \tilde qH_aH_c \cr
(\bar E_2,E_2) &= \tilde qH_aH_b \cr
(\bar E_3,E_3) &= \tilde qH_bH_c \cr
& \hfil \cr 
\tilde q &= q-q_1 \cr} \quad
\eqalign{{\rm b)}~&(\bar E_1,H_b) = \tilde q\bar E_2\bar E_3 \cr
&(\bar E_1,E_2) = \tilde qH_a\bar E_3 \cr
&(\bar E_3,E_1) = \tilde qH_cE_2 \cr
& \hfil \cr 
& \hfil \cr} \quad
\eqalign{{\rm c)}~&\bar E_2E_3 = qE_3\bar E_2 \cr
&E_2E_1 = qE_1E_2 \cr &E_3E_1 = qE_1E_3 \cr
&E_2E_3 = E_3E_2 \cr
& \hfil \cr}  \eqno(10.4) 
$$

Let us introduce the following basis states:
$$
|n_1^in_2^in_3^i\rangle = \bar E_1^{n_1^i}\bar E_2^{n_2^i}\bar E_3^{n_3^i}|
000\rangle \qquad i = a,b,c \eqno(10.5)
$$
\no where the $\bar E_i$ are regarded as raising operators.  Then
$$
\eqalignno{H_a|n_1^an_2^an_3^a\rangle =
q_1^{n_1^a+n_2^a}\alpha_a|n_1^an_2^an_3^a\rangle \qquad
&n_3^a = 0 & (10.6) \cr
H_b|n_1^bn_2^bn_3^b\rangle = q_1^{n_2^b+n_3^b}\alpha_b|n_1^bn_2^bn_3^b\rangle
\qquad & n_1^b = 0 & (10.7) \cr
H_c|n_1^cn_2^cn_3^c\rangle = q_1^{n_1^c+n_3^c}\alpha_c|n_1^cn_2^cn_3^c\rangle
\qquad & n_2^c = 0 & (10.8) \cr}
$$
\no These eigenstates are restricted by the requirement that $H_i$ and the
associated $E_j$ lie in the same row or column. e.g. $H_a$ lies in the same
column as $E_2$ and $E_1$.

Let the space of states corresponding to $H_i$ be ${\cal{H}}_i$~
$(i=a,b,c)$.  Let the states belonging to ${\cal{H}}_a\otimes
{\cal{H}}_b \otimes {\cal{H}}_c$ be denoted by
$$
\eqalign{&|n_1^an_2^an_3^a;~n_1^bn_2^bn_3^b;~n_1^cn_2^cn_3^c\rangle 
\quad \hbox{or} \quad \cr
&|\vec n^a\vec n^b\vec n^c\rangle \quad \hbox{or simply} \quad
|n\rangle~.  \cr} \eqno(10.9)
$$

The matrix elements of $E_i$ and $\bar E_i$ in this basis are restricted
by selection rules following from (10.3).  For example
$$
\eqalign{&E_2H_a = q_1H_aE_2 \cr
\langle n^\prime|&E_2H_a-q_1H_aE_2|n\rangle = 0 \cr} \eqno(10.10)
$$
\no in the notation of (10.9).  By (10.6)
$$
(q_1^{n_1^{a'}+n_2^{a'}+1}-q_1^{n_1^a+n_2^a})\langle n^\prime|E_2|n\rangle = 0~.
$$
\no Then
$$
\langle n^\prime|E_2|n\rangle = 0 \quad \hbox{if} \quad
n_1^{a'}+n_2^{a'}\not= n_1^a+n_2^a-1~. \eqno(10.11)
$$
\no We also have
$$
\langle n^\prime|E_2H_b-q_1H_bE_2|n\rangle = 0 \eqno(10.12)
$$
\no implying
$$
\langle n^\prime|E_2|n\rangle = 0 \qquad \hbox{if} \qquad
n_2^{b'}+n_3^{b'}\not= n_2^b+n_3^b-1 \eqno(10.13)
$$
\no In general we have
$$
\langle n-\Delta_i|E_i|n\rangle = 0 \qquad i = 1,2,3 \eqno(10.14)
$$
\no unless $\Delta_i$ is one of the following four possibilities:
$$
\eqalign{&\underline{\Delta_1} \cr
&~~{\rm a}~~~~~~{\rm b}~~~~~~{\rm c} \cr
&(100;~000;~100) \cr &(100;~000;~001) \cr
&(010;~000;~100) \cr &(010;~000;~001) \cr} \qquad
\eqalign{&\underline{\Delta_2} \cr
&~~{\rm a}~~~~~~{\rm b}~~~~~~{\rm c} \cr
&(100;~010;~000) \cr &(010;~010;~000) \cr
&(100;~001;~000) \cr &(010;~001;~000) \cr} \qquad
\eqalign{&\underline{\Delta_3} \cr
&~~{\rm a}~~~~~~{\rm b}~~~~~~{\rm c} \cr
&(000;~010;~100) \cr &(000;~001;~100) \cr
&(000;~010;~001) \cr &(000;~001;~001) \cr} \eqno(10.15)
$$
\no where the entries in this table represent
$$
\Delta_i(n_1^an_2^an_3^a;~n_1^bn_2^bn_3^b;~n_1^cn_2^cn_3^c) \qquad
i=1,2,3 
$$
\no  Among the equations of (10.4b) there are relations such as
$$
(H_b,E_1) = \tilde q E_2E_3 \eqno(10.16)
$$
\no implying by (10.7) and (10.14)
$$
\alpha_b(q_1^{n_2^{b'}+n_3^{b'}}-q_1^{n_2^b+n_3^b})
\langle n^\prime|E_1|n\rangle = \tilde q \sum_{\Delta_3}
\langle n^\prime|E_2|n-\Delta_3\rangle\langle n-\Delta_3|E_3|n\rangle~.
\eqno(10.17) 
$$

By the selection rules on $E_2$ and $E_3$ the right side of the preceding
equation vanishes unless $n^\prime=n-\Delta_2-\Delta_3$; but the left
side vanishes unless $n^\prime = n-\Delta_1$ by the selection rules on
$E_1$.  Since $\Delta_1\not= \Delta_2+\Delta_3$ by (10.15), it follows
that
$$
\eqalignno{\sum_{\Delta_3}&\langle n-\Delta_2-\Delta_3|E_2|n-\Delta_3\rangle\langle n-\Delta_3|E_3|n\rangle
= 0 & (10.18) \cr
\noalign{\hbox{or}}
&\langle n^\prime|E_2E_3|n\rangle = 0 & (10.19a) \cr
\noalign{\hbox{and by (10.16)}}
&\langle n^\prime|(H_b,E_1)|n\rangle = 0 & (10.19b) \cr}
$$
\no Similar remarks hold for the other two equations of (10.4b), i.e.
$$
\eqalignno{&\langle n^\prime|(\bar E_1,E_2)|n\rangle = 0 \quad
\hbox{since} \quad -\Delta_1 + \Delta_2 \not= -\Delta_3 & (10.20) \cr
&\langle n^\prime|(\bar E_3,E_1)|n\rangle = 0 \quad
\hbox{since} \quad -\Delta_3+\Delta_1 \not= \Delta_2 & (10.21) \cr}
$$

\ve

\line{{\bf XI. The Amplitudes of the Raising and Lowering Operators.} \hfil}
\vskip.3cm

The $E_i$ and their hermitian conjugates are restricted by the following
relations (10.4a)
$$
\eqalign{(\bar E_1,E_1) &= \tilde q H_aH_c \cr
(\bar E_2,E_2) &= \tilde qH_aH_b \cr
(\bar E_3,E_3) &= \tilde q H_bH_c \cr} \eqno(11.1)
$$
\no Consider for example
$$
\langle n|\bar E_2E_2-E_2\bar E_2|n\rangle = \tilde q
\langle n|H_aH_b|n\rangle \eqno(11.2)
$$
\no or
$$
\sum_p\langle n|\bar E_2|p\rangle\langle p|E_2|n\rangle -
\sum_{p^\prime}\langle n|E_2|p^\prime\rangle\langle p^\prime|\bar E_2|n\rangle
= \tilde q\langle n|H_aH_b|n\rangle
$$
\no in the notation of (10.9) for $|n\rangle$.

Since $E_2$ and $\bar E_2$ are hermitian conjugate, the preceding
equation may be written as follows:
$$
\sum_p|\langle p|E_2|n\rangle|^2-\sum_{p^\prime}\langle n|E_2|
p^\prime\rangle|^2 = \tilde q\langle n|H_aH_b|n\rangle \eqno(11.3)
$$
\no or
$$
\sum_{\Delta_2}|\langle n-\Delta_2|E_2|n\rangle|^2 - \sum_{\Delta_2}
|\langle n|E_2|n+\Delta_2\rangle|^2 = \tilde q\alpha_a
\alpha_b q_1^{n_1^a+n_2^a+n_2^b+n_3^b} \eqno(11.4)
$$
\no where $\Delta_2$ is summed over the 4 possibilities shown in (10.15).

There are corresponding equations for $E_1$ and $E_3$.

In addition to the equations (10.4a) there are the equations
(10.4b) and (10.4c) as well as (10.3) that the three $E_i$ matrices
must satisfy.

Note that if $|n\rangle$ is a ground state, the first term in (11.4)
vanishes.  Then (11.4) implies
$$
\tilde q\alpha_a\alpha_b < 0~. \eqno(11.5)
$$
\no There are similar equations for $E_1$ and $E_3$ and therefore
$$
\tilde q\alpha_i\alpha_j < 0 \eqno(11.6)
$$
\no for $(i,j) = (a,b,c)$.  Hence the $\alpha_i$ are all of the same
sign; therefore $\alpha_i\alpha_j$ are positive and
$$
\tilde q = q-q_1 < 0 \eqno(11.7)
$$
\no so that
$$
q<1~. \eqno(11.8)
$$

One may regard the three $H_i$ operators as the formal Hamiltonians of
three $q$-\break oscillators and the three ${\cal{H}}_i$ as their state spaces.
One may also regard these same oscillators as field oscillators associated
with three distinct fields.  Then the $E_i$ and $\bar E_i$ correspond to
absorption and emission operators.  In the same language the $E_i$ and
$\bar E_i$ describe associated absorption and emission of two particles,
since according to (10.15) two population numbers change in every case.
\vskip.5cm

\line{{\bf XII. The Quantum Determinant of $SL_q(3)$.} \hfil}
\vskip.3cm

Let
$$
\Delta = {\rm det}_q T \eqno(12.1)
$$
\no Then by (9.8)
$$
\eqalign{\Delta = \bar E_1H_bE_1&-q[\bar E_2\bar E_3E_1 +
\bar E_1E_2E_3] \cr
&-q^3H_aH_bH_c + q^2[\bar E_2E_2H_c + H_a\bar E_3E_3]~. \cr} \eqno(12.2)
$$
\no Since $\Delta$ belongs to the center of the algebra, we set
$$
\Delta = 1 \eqno(12.3)
$$
\no Applied to the $H$-vacuum we have
$$
\Delta|0\rangle = |0\rangle \eqno(12.4)
$$
\no By (12.2), (12.3) and (10.6)-(10.8)
$$
\alpha_a\alpha_b\alpha_c = -q_1^3 \eqno(12.5)
$$
\no We also have
$$
\langle\vec n|\Delta|\vec p\rangle = \langle\vec n|\Delta|\vec n\rangle
\delta(\vec n,\vec p) \eqno(12.6)
$$
\no and
$$
\langle n|\Delta|n\rangle = 1 \eqno(12.7)
$$
\no By (12.2), (10.3) and (10.4)
$$
\eqalign{\langle n|\Delta|n\rangle = &H_b(n)
\langle n|\bar E_1E_1|n\rangle - q^3\langle n|H_aH_bH_c|n\rangle \cr
&+q^2[H_c(n)\langle n|\bar E_2E_2|n\rangle +
H_a(n)\langle n|\bar E_3E_3|n\rangle] \cr} \eqno(12.8)
$$
\no or by (12.5) and (12.7)
$$
\eqalign{1-q_1^{n^a+n^b+n^c} = \alpha_bq_1^{n_b}
\langle n|\bar E_1E_1|n\rangle &+ q^2\alpha_cq_1^{n_c}
\langle n|\bar E_2E_2|n\rangle \cr
&+q^2\alpha_aq_1^{n_a}\langle n|\bar E_3E_3|n\rangle \cr} \eqno(12.9)
$$
\no where
$$
\eqalign{n^a &= n_1^a + n_2^a \cr n^b &= n_2^b + n_3^b \cr
n^c &= n_1^c + n_3^c \cr} \eqno(12.10)
$$
\no Eq. (12.9) is consistent with our earlier conclusion that the
$\alpha_i$ are negative and $q_1>1$.
\vskip.5cm

\line{{\bf XIII. The Dual Algebras.} \hfil}
\vskip.3cm

We have already described the $q$-algebra dual to $SU_q(2)$.  The
corresponding $q$-algebra dual to the higher $q$-groups is described by
the following equations as expressed in the Chevalley-Serre basis$^{8,9,10}$
$$
\eqalignno{(H_i,H_j) &= 0 & (13.1) \cr
(H_i,X_j^\pm) &= \pm(\alpha_i,\alpha_j) X_j^\pm 
\qquad i,j = 1~\ldots~ r & (13.2) \cr
(X_i^+,X_j^-) &= \delta_{ij}{\sinh~h~H_i\over \sinh~h} = \delta_{ij}
{q^{H_i}-q_1^{H_i}\over q-q_1} & (13.3a) \cr}
$$
\no where $r$ is the rank and
$$
q = e^h \eqno(13.3b)
$$

The Serre relations are
$$
\sum^m_{k=0} (-)^k \left(\matrix{m \cr k \cr} \right)_q
\hat q_i^{-k(m-k)/2} (X_i^\pm)^k X_j^\pm (X_i^\pm)^{m-k} = 0 \qquad
i\not= j \eqno(13.4a)
$$
\no where
$$
\eqalign{m &= 1-A_{ij} \cr
\hat q_i &= e^{h(\alpha_i,\alpha_i)} \cr
\left(\matrix{m \cr k \cr} \right)_q &= {\langle m\rangle_q!\over
\langle k\rangle_q!\langle m-k\rangle_q!} \cr} \eqno(13.4b)
$$
\no Here the $\alpha_i$ are the simple roots and $A_{ij}$ are the
elements of the Cartan matrix for the simple Lie algebras.
\vskip.5cm

\line{{\bf XIV. Invariant Bilinears and Lagrangians of a Toy Theory.} \hfil}
\vskip.3cm

Eqs. (9.1) may be rewritten as
$$
(T_2T_1) R(T_2^{-1}T_1^{-1}) = (T_1^{-1}T_2^{-1}) R(T_1T_2) = R \eqno(14.1)
$$
\no Set
$$
\eqalignno{{\cal{T}} &= T_1T_2 & (14.2) \cr
\hat{\cal{T}} &= T_2T_1 & (14.3) \cr}
$$
\no ${\cal{T}}$ lies in two spaces which are permuted in $\hat{\cal{T}}$.  
Then
$$
\hat{\cal{T}}R{\cal{T}}^{-1} = \hat{\cal{T}}^{-1}R{\cal{T}} = R \eqno(14.4)
$$
\no When $N=2$, one also has (1.1):
$$
T\epsilon T^t = T^t\epsilon T = \epsilon \eqno(14.5)
$$
\no where $\epsilon$ is given by (1.2) while
$$
R = \left(\matrix{q & 0 & 0 & 0 \cr
0 & 1 & 0 & 0 \cr
0 & \tilde q & 1 & 0 \cr
0 & 0 & 0 & q \cr} \right) \eqno(14.6)
$$

Let $\psi$ and $\hat\psi$ be matrix fields transforming as
$$
\eqalign{\psi^\prime &= {\cal{T}}\psi \cr
\hat\psi^\prime &= \hat\psi\hat{\cal{T}}^{-1} \cr} \quad \hbox{or} \quad
\eqalign{\psi^\prime_{\ell k} &= \sum (T_1)_{\ell s} (T_2)_{kt}
\psi_{st} \cr
\hat\psi^\prime_{pm} &= \sum\hat\psi_{ji}(T_1^{-1})_{jp}(T_2^{-1})_{im} \cr}
\eqno(14.7)
$$
\no Then by (14.4)
$$
(\hat\psi R\psi)^\prime = \hat\psi R\psi \eqno(14.8)
$$
\no is the bilinear invariant corresponding to $\psi^t\epsilon\psi$.  

To construct a toy theory one may distinguish between $H$ and $E$
fields by associating the $H$ with a Lorentz spinor field and
$E$ with a Lorentz vector field in a particular gauge.

This association leads to a supersymmetric feature since the spinor and vector
then belong to the same multiplet and therefore transform into each other
under a $q$-\break transformation. In this special gauge one may write the spinor field as
$$
\psi_{ij} = \sum_k\varphi^k_{ij}H(k) \qquad k = a,b,c \eqno(14.9)
$$
\no and a mass term as
$$
M\hat\psi CR\psi = M\sum_k H(k)^2 U(k)  \eqno(14.10)
$$
\no where $\psi_{ij}$ transforms as (14.7) and we orthonormalize as follows
$$
\hat\varphi^j CR\varphi^k = \delta^{jk} U(k) \eqno(14.11)
$$
\no With respect to the Dirac algebra $\hat\varphi^j$ is the transposed spinor
and $C$ is the charge conjugation matrix.

The eigenvalues of the mass term in the state $|n^an^bn^c\rangle$
are
$$
M\sum_k q_1^{2n^k}\alpha_k^2 U(k) \qquad k = a,b,c \eqno(14.12)
$$
\no where by (12.5)
$$
\alpha_a\alpha_b\alpha_c = -q_1^3~. \eqno(14.13)
$$
\no One may compare (14.12) and (14.13) with (4.9) and (1.11).  In
(14.12) there is a rising rather than an inverted mass spectrum since
$q$ is there replaced by $q_1$ and in both cases $q<1$.  (The difference
between the two cases comes from an arbitrary switch of the positions
of the Hermitian conjugate operators with respect to the
minor diagonal of $T$.)

Both spectra and also the spectrum (5.8) imply an underlying soliton
structure with constituent fields associated with the elements of the
$q$-algebra.  These three spinor fields may interact with the three
vector fields through terms like
$$
\hat\psi CR\nabla\!\!\!\!/\psi \eqno(14.14)
$$
\no where
$$
\nabla_\mu = \partial_\mu + A_\mu~, \qquad 
\nabla\!\!\!\!/ = \gamma^\mu\nabla_\mu~. \eqno(14.15)
$$
\no Here the vector field is
$$
A\!\!\!/ = \sum^3_{\alpha=1} \bigl(A\!\!\!/(\alpha) E(\alpha) +
\bar A\!\!\!/(\alpha)\bar E(\alpha)\bigr)~. \eqno(14.16)
$$
\no The $q$-invariance of the interaction term requires the following
$q$-transformation laws for the covariant derivative
$$
\nabla\!\!\!\!/~^\prime = {\cal{T}}\nabla\!\!\!\!/{\cal{T}}^{-1}
\eqno(14.17)
$$
\no and the $q$-vector connection
$$
A\!\!\!/^\prime = {\cal{T}}A\!\!\!/{\cal{T}}^{-1} +
{\cal{T}}\partial\!\!\!/{\cal{T}}^{-1}~. \eqno(14.18)
$$
\no The general relations shown in Section 6 will continue to hold
here if ${\cal{T}}$ is substituted for $T$.  In particular the form (6.8)
will replace the usual trace
$$
{\rm Tr}~ F_{\mu\nu}F^{\mu\nu}
$$
\no even though
$$
F_{\mu\nu}^\prime = {\cal{T}} F_{\mu\nu}{\cal{T}}^{-1}
$$
\no because the matrix elements of $F_{\mu\nu}$ lie in the $q$-algebra.

The preceding discussion can be generalized to any $N$, including $N=2$,
by representing the fermion field as
$$
\psi = \sum^N_1 \varphi^i H_i~. \eqno(14.19)
$$
\no In the $N=2$ case we have put $H_1 = \beta$, $H_2 = \gamma$ in
Section 3.  The interacting boson field may be written as
$$
A\!\!\!/ = \sum_{i<j} (\bar A_{ij}\bar E_{ij} + A\!\!\!/_{ji}E_{ji}) \qquad
i,j = 1\ldots N \eqno(14.20)
$$
\no The general fermionic mass term is
$$
M\hat\psi CR\psi = M\sum^N_{k=1} q_1^{2n^k}\alpha_k^2 U(k) \eqno(14.21)
$$
\no where we have again normalized as follows
$$
\hat\varphi^kCR\varphi^j = \delta^{kj} U(k)~. \eqno(14.22)
$$
\no Here, by the argument leading to (14.13)
$$
\prod^N_i \alpha_k = -q_1^N~. \eqno(14.23)
$$
\no The lowest order interaction term is
$$
\hat\psi CRA\!\!\!/ \psi~. \eqno(14.24)
$$
\no The matrix element between arbitrary states is
$$
\eqalignno{\langle n^\prime&|\bigl(\sum_k\hat\varphi_kH_k\bigr) CR
\bigl(\sum_{j>i} A\!\!\!/_{ij}E_{ij}\bigr)
\bigl(\sum_\ell\varphi_\ell H_\ell\bigr)|n\rangle +
\sum_{i<j}\bar A\!\!\!/_{ij}\bar E_{ij}~\hbox{terms} & (14.25) \cr
&= \sum_{j>i}\sum_{k,\ell}(\hat\varphi_k CRA\!\!\!/_{ij}\varphi_\ell)
\langle n^\prime|H_kE_{ij}H_\ell|n\rangle + \sum_{i<j} \ldots \cr
&= \sum_{j>i}\sum_{k,\ell}(\hat\varphi_k CRA\!\!\!/_{ij}\varphi_\ell)
q_1^{n_k}q_1^{n_\ell}\alpha_k\alpha_\ell
\langle n^\prime|E_{ij}|n\rangle + \sum_{i<j} \quad \ldots & (14.26) \cr}
$$
\no i.e., the $q$-Lorentz invariant form $\hat\varphi_k CRA\!\!\!/_{ij}
\varphi_\ell$ is averaged over the algebra in the way shown.

In this way and in other procedures that generalize naturally from usual
field theory one could in principle construct a quantum mechanical
description of the interaction of these constituent fields.

Finally we return to the
picture in which the particles of the theory are solitons composed of
the constituent fields.  We shall work at the level of $SU_q(2)$.  In this
picture the fields may be represented
by the expansions (5.1).
Then the mass of a single soliton depends on expressions like (5.5) and
$$
\ldots h\bigl[(\tilde\varphi^j_{mp}\bar{\cal{D}}^j_{mp})
(\varphi^j_{mp}{\cal{D}}^j_{mp})\bigr] \eqno(14.27)
$$
\no which leads to (5.8).  

Similarly the lowest order interaction term in the same model is
$$
\ldots~h\biggl[\bigl((\psi^{j_1}_{m_1p_1})^t\bar{\cal{D}}^{j_1}_{m_1p_1}\bigr)
\epsilon C(A\!\!\!/^{j_2}_{m_2p_2}{\cal{D}}^{j_2}_{m_2p_2})
(\psi^{j_3}_{m_3p_3}{\cal{D}}^{j_3}_{m_3p_3})\biggr] \eqno(14.28)
$$
\no The preceding
expression reduces to the following:
$$
\eqalign{&\bigl[\bigl(\psi^t)^{j_1}_{m_1p_1}\epsilon CA\!\!\!/^{j_2}_{m_2p_2}\psi^{j_3}_{m_3p_3}\bigr] h\bigl(\bar{\cal{D}}^{j_1}_{m_1p_1}
{\cal{D}}^{j_2}_{m_2p_2}{\cal{D}}^{j_3}_{m_3p_3}\bigr) \cr
&=\bigl[(\psi^t)^{j_1}_{m_1p_1}\epsilon CA\!\!\!/^{j_2}_{m_2p_2}
\psi^{j_3}_{m_3p_3}\bigr]
\left[\matrix{j_1j_2j_3 \cr m_1m_2m_3 \cr}\right]_q~~
\left[\matrix{j_1j_2j_3 \cr p_1p_2p_3 \cr}\right]_q \cr} \eqno(14.29)
$$
\no in terms of $q$-Clebsch-Gordan coefficients$^{11}$

In the limit $q=1$, there is no internal structure $(j_1=j_2=j_3=0)$ and (14.29)
reduces to
$$
\psi^tCA\!\!\!/\psi \eqno(14.30)
$$

\vskip.5cm

\line{{\bf XV. Summary.} \hfil}
\vskip.3cm

There are two $q$-algebras, $G_q$ and $g_q$, that are respective
deformations of the Lie group $(G)$ and its algebra $(g)$.

In a matrix representation the matrix elements of $G,g$, and $g_q$ all
commute, but the matrix elements of $G_q$ form a non-commuting algebra.

One ordinarily associates a vector connection with the Lie algebra of
$SU(2)$ to obtain the electroweak vectors and similarly one associates
a vector connection with the Lie algebra of $SU(3)$ to obtain the gluons
of the standard model.

If one proceeds in the same way with the deformed Lie algebras $g_q$
of $SU_q(2)$ and $SU_q(3)$ one obtains a hypothetical theory lying close
to the standard theory but with differences that can in principle be
computed by a perturbation expansion in $q$.  In the limit $q=1$ one
ought to recover the standard theory and when $q$ is near unity, the
$q$-theory may have the good formal properties of the standard theory.
Here, however, we have only succeeded in discussing a toy theory based on
$SL_q(3)$ and not on $SU_q(3)$.

In the hypothetical $q$-theory there is also the dual algebra $G_q$ with
which one may also associate a different set of hypothetical dual fields that
play the role of quark-like and gluon-like constituent fields.  

The two sets of fields, associated with the two dual algebras, do not represent independent sectors of the full theory.  They should rather be regarded as
complementary descriptions; one picture is microscopic and the second
picture, approximating the standard theory, is phenomenological.

Since the Lie groups used in particle physics are, unlike the Poincar\'e
group, only phenomenological, they have no greater {\it a priori} claim
than the $q$-groups.  Therefore the $q$-theories must be judged, just as
the Lie theories, by their phenomenological usefulness.
\vskip.5cm

\line{{\bf References.} \hfil}
\vskip.3cm

\item{1.} See, for example, {\it Quantum Groups}, eds. T. Cartwright,
D. Fairlee, and C. Zachos (World Scientific, 1991); {\it Yang-Baxter
Equation in Integrable Systems}, M. Jimbo (World Scientific, 1989);
{\it Affine Lie Algebras and Quantum Groups}, ed. Jurgen Fuchs (Cambridge,
1992).

\item{2.} R. J. Finkelstein, Mod. Phys. Lett. A{\bf 15}, 1709 (2000).

\item{3.} S. L. Woronowicz, RIMS Kyoto {\bf 23}, 112 (1987).

\item{4.} R. J. Finkelstein, Int. J. Mod. Phys. {\bf 11}, 733 (1996).

\item{5.} R. J. Finkelstein, Lett. Math. Phys. {\bf 54}, 157 (2000).

\item{6.} R. J. Finkelstein, E. Marcus, J. Math. Phys. {\bf 36}, 2652
(1995).

\item{7.} R. J. Finkelstein, J. Math. Phys. {\bf 37}, 2628 (1996).

\item{8.} N. Yu. Reshetikhin, L. A. Takhtadzhan, and L. D. Fadeev,
Leningrad Math. J{\bf 1}, 191 (1990).

\item{9.} V. G. Drinfeld, Dokl. Akad. Nauk. SSSR {\bf 283} (1985).

\item{10.} M. Jimbo, Lett. Math. Phys. {\bf 11}, 247 (1986).

\item{11.} See, for example, A. C. Cadavid and R. J. Finkelstein, J. Math. Phys. {\bf 36}, 1912 (1995).

\end
\bye